\documentclass[aps,prx,twocolumn,longbibliography,superscriptaddress,showpacs]{revtex4-1}
\usepackage{graphicx}
\usepackage{latexsym}
\usepackage{amssymb}
\usepackage{amsmath}
\usepackage{amsfonts}
\usepackage{mathtools}
\usepackage{dsfont}
\usepackage{bbm}
\usepackage{bm}
\usepackage{multirow}
\usepackage{color}
\usepackage{tikz}
\usepackage{comment}
\usepackage{footmisc} 
\usepackage{xcolor}

\usepackage[colorlinks=true, citecolor={blue!80!black}, urlcolor={blue!50!black}, linkcolor = {blue!80!black}]{hyperref}
\allowdisplaybreaks[1]
\usepackage[percent]{overpic}
 \usepackage[utf8]{inputenc}

\DeclareMathAlphabet{\mathbbold}{U}{bbold}{m}{n}

\definecolor{forestgreen}{rgb}{0.13, 0.55, 0.13}
\definecolor{gr}{rgb}{0,0.82,0.18}
\definecolor{DarkerRed}{rgb}{0.9,0.05,0.25}

\begin{document}
\title{Universal crossovers in weakly-monitored quantum critical states}

\author{Abhishek Kumar}
\email{akumar0@umass.edu}

\affiliation{Department of Physics, University of Massachusetts, Amherst, Massachusetts 01003, USA}
\author{Rushikesh~A.~Patil}
\affiliation{Department of Physics, University of California, Santa Barbara, California 93106, USA}
\author{Andreas~W.~W.~Ludwig}
\affiliation{Department of Physics, University of California, Santa Barbara, California 93106, USA}
\author{Romain Vasseur}
\affiliation{Department of Theoretical Physics, University of Geneva,
  24 quai Ernest-Ansermet, 1211 Gen\`eve, Switzerland}
\affiliation{Geneva Quantum Center, University of Geneva}

\date{\today}

\begin{abstract}

    We study post-measurement ensembles of  ground states of tricritical and critical 1D quantum Ising Hamiltonians
    subjected, respectively, to weak energy and spin measurements without post\-selection. These measurements act as relevant perturbations about the unmeasured critical ground states. Using finite-size renormalization group (RG) crossover analyses, we characterize their universal properties through the 
    entanglement
    effective central charge, effective Affleck-Ludwig  boundary  entropy, and 
    signatures of
    multifractality from
    moments of measurement-averaged  correlation functions.  In both cases, we find evidence for ``measurement-dominated'' or ``measurement-altered'' fixed points governed by the underlying Born-rule randomness. For critical Ising, we find a direct RG flow to a projective-measurement fixed point with area-law entanglement, whereas for the tricritical Ising model, we find evidence for a 
    weak-measurement 
    fixed point with 
    logarithmic
    entanglement. These results clarify the RG-flow structure of weakly measured multicritical Ising ground states and show how intrinsic measurement-induced randomness can generate complex and rich universal long-distance  scaling  behavior in the post-measurement ensembles, accessible to controlled analytical RG and numerical finite-size RG crossover analyses.

\end{abstract}

\maketitle

\section{Introduction}

\begin{figure*}[t!]
	\centering
 \includegraphics[width=\textwidth]{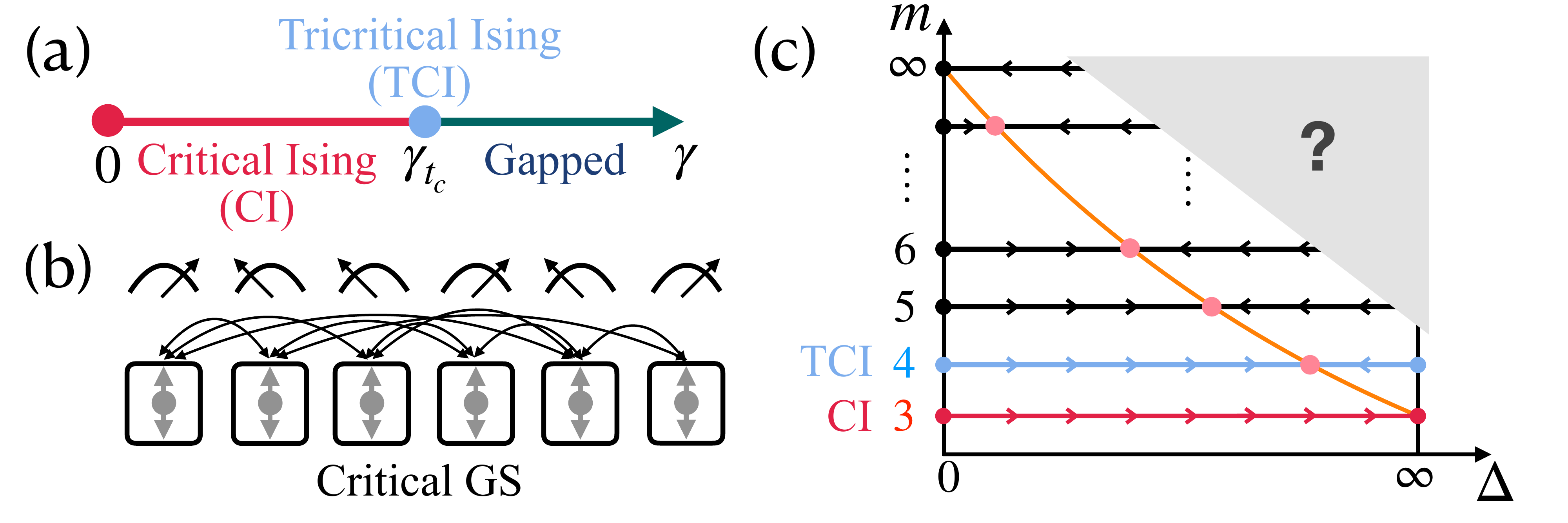}
    \caption{\textbf{Model setup and RG flow.}~(a) Zero-temperature phase diagram of the one-dimensional O'Brien--Fendley chain. The three-spin interaction controlled by $\gamma$ drives the critical Ising (CI) point, shown in red, to a gapped phase through the tricritical Ising (TCI) point at $\gamma_{tc}$, shown in sky blue.~(b) Schematic illustration of weak measurement operations applied to the critical ground states.~(c) RG-flow schematic for weakly measured critical ground states. The CI and TCI points correspond to the $m=3$ and $m=4$ fixed-point actions of the multi-critical Ising Landau-Ginzburg sequence, which are famously known as the $m^{\rm th}$ unitary minimal model CFT. Weak measurements with the operator considered in Eq.~\ref{replica_action_full}  
    act as a relevant perturbation in the RG sense
    to the unmeasured critical ground state. For CI, $m=3$, the flow runs directly toward the projective-measurement fixed point, $\Delta\to\infty$. For TCI, $m=4$, our numerics support an infrared weak-measurement fixed point, at finite effective measurement strength $\Delta$.
    Using the fact that measurements are marginally irrelevant at $m = \infty$, a 
    weak-measurement
    fixed point occurring at $\mathcal{O}(\epsilon)$, with $\epsilon=\frac{3}{m+1}$ for large $m$, was demonstrated in 
    Ref.~\cite{PatilLudwig2024}.
    This leads to the conjecture of the phase diagram with 
    a weak-measurement
    infrared fixed point 
    for finite $m>3$ (indicated by the orange dots), whereas at $m=3$ the weak-measurement and the projective-measurement fixed point become one and the same, as illustrated in the figure. Note that in addition to the 
    weak-measurement
    fixed point 
    shown in orange, it might be possible to have more fixed points which occur at larger measurement strengths in the shaded gray region, and their existence will likely depend on the explicit lattice critical ground state and also the microscopic measurement protocol. Since we only consider explicit lattice measurement protocols for the critical Ising and the tricritical Ising cases, we leave the latter investigation for higher multicritical Ising ground states under measurements to a future work.\\
}
\label{fig:model_setup}
\end{figure*}

A quantum measurement is intrinsically non-unitary and stochastic, each measurement record 
 preparing a different outcome-conditioned wavefunction, with probabilities determined by the Born rule. Thus, even for  a  disorder-free unmeasured  many-body quantum system, the ensemble of post-measurement states carries intrinsic 
`quantum mechanical' randomness. More broadly, measurement-conditioned ensembles have emerged as a useful framework for probing many-body quantum states at the level of wavefunction ensembles. In {\it projected ensembles}, measurement on part of a many-body wavefunction defines an ensemble of conditional states on the remaining subsystem, which can approach universal state-design distributions -- a phenomenon known as deep thermalization~\cite{PhysRevLett.128.060601,PRXQuantum.4.010311,PRXQuantum.4.030322}. Closely related learning and inference perspectives have further shown that conditioning on measurement records, or learning partial information about an initially unknown configuration, can itself generate sharp transitions and novel RG fixed points~\cite{PhysRevLett.129.120604,PhysRevX.12.041002,PhysRevLett.129.200602,PhysRevLett.130.220404,PRXQuantum.5.020304,PhysRevB.109.094209,7dpt-d4s5,295c-lj1w,4dwm-kn11,b3s8-y1wb,kim2025learning,patel2026universal,patil2026highernishimori,WieseDasNahum}.  These developments highlight how conditioning on measurement outcomes can generate universal structures at the level of wavefunction ensembles and probability distributions. Understanding how this randomness modifies universal  critical properties is therefore a central question in measurement-induced phenomena. 

This question has been explored most extensively in monitored quantum circuits, where measurements compete with unitary dynamics and can drive measurement-induced entanglement transitions between volume-law and area-law phases~\cite{PhysRevB.98.205136,PhysRevB.99.224307,PhysRevX.9.031009,PhysRevB.100.134306,potter2022entanglement,Fisher:2022qey,PhysRevLett.125.030505,PhysRevB.103.104306,PhysRevB.103.174309,PhysRevX.10.041020,Li_2024}, and in related transitions occurring in circuits of non-interacting fermions (e.g., Refs.~\cite{jian2023shapourian,FavaNahum2023,yang2026coherent}). At the transition, the entanglement 
exhibits logarithmic scaling, where the critical behavior 
is characterized by a non-unitary 
(and logarithmic) conformal field theory (CFT)~\cite{PhysRevB.101.060301, PhysRevB.101.104301, PhysRevB.101.104302, PhysRevB.104.104305, PhysRevLett.125.070606, PhysRevLett.128.050602, PhysRevB.109.014303}. These developments show that measurements can generate  entirely novel universality classes that are qualitatively different from those occurring in closed, unitary quantum systems. The key distinguishing feature of the former from the latter is the special intrinsic randomness of quantum mechanical measurement outcomes, which is the source of the non-unitarity of the resulting critical phenomena. 

A complementary setting 
to investigate universal effects of measurements
is provided by 
starting with
equilibrium critical ground states and subjecting them to measurements. In this case, the unmeasured state already has a controlled long-distance description in terms of 
a (unitary)
conformal field theory, and a layer of measurements acts as a random perturbation localized on the zero imaginary-time slice of the corresponding Euclidean spacetime geometry. This viewpoint, initiated in~\cite{PhysRevX.13.021026}, showed that measurements acting on Luttinger-liquid ground states imply a random defect perturbation on the measurement time-slice which induces a Kosterlitz-Thouless type defect phase transition.
Related studies of critical Ising chains  and similar systems
~\cite{PhysRevB.107.245132,PhysRevB.108.165120,
PhysRevX.13.041042,sun2023new,l4b7-h5cd, Naus:2025ivu,chen2025zippingmanybodyquantumstates}
considered local measurements with and without postselection~\footnote{Spatially uniform postselection of measurement outcomes leads to standard and well-studied unitary boundary critical phenomena~\cite{CARDY1989Fusion,AffleckLudwigNPB1991,LudwigIJMPB1994,affleck1995ActaPhysPol}, because it removes the randomness of the measurement outcomes, which is the source of non-unitarity.}, while other works~\cite{PhysRevLett.130.250403,LeeJianXu,PhysRevB.110.094404,PhysRevLett.134.096503} investigated decoherence and measurement channels acting on one-dimensional critical states. Together, these results establish single-shot measurements of critical ground states as a controlled arena for studying measurement-generated universality beyond monitored circuit dynamics. 

Of particular relevance to the present work is the 
{\it controlled RG analysis} of 
measurement problems on one-dimensional
quantum critical ground states without postselection, that was developed in Ref.~\cite{PatilLudwig2024} (see also~\cite{JengLudwig}). 
The two central examples were the tricritical Ising ground state subjected to weak measurements of the local energy operator and the critical Ising ground state subjected to weak measurements of the local spin operator. In both cases,
 measurements induce a perturbation that is  RG-relevant at the 
unmeasured fixed point, leading to 
rich and complex criticality 
at a \textit{measurement-dominated fixed point}, which was found in the epsilon expansion analysis in Ref.~\cite{PatilLudwig2024}. 
This measurement-dominated fixed point controls the long-distance behavior of the post-measurement ensemble, and the epsilon expansion analysis demonstrates unconventional universal properties of this fixed point, including a hierarchy of entanglement effective central charges, effective Affleck-Ludwig boundary entropy manifested through the Shannon entropy of the measurement record,
and multifractal scaling dimensions for 
moments of correlation functions. These properties reflect the \textit{non-unitary} nature of the boundary critical phenomenon described by the {\it measurement-dominated} fixed point arising from Born-rule randomness. They would be replaced by  properties of standard and conventional {\it unitary} boundary critical phenomena~\cite{CARDY1989Fusion,AffleckLudwigNPB1991,LudwigIJMPB1994,affleck1995ActaPhysPol} if the measurement outcomes were to be uniformly postselected (as observed, e.g., in~\cite{l4b7-h5cd}).

In this work, we numerically test and extend this RG 
analysis
using microscopic lattice realizations of critical Ising and tricritical Ising ground states, subjected respectively to single-shot local spin and local energy measurements. First, we characterize the universal properties of the post-measurement ensembles associated with two distinct Ising criticalities governed by different minimal-model CFTs: the $m=3$ critical Ising CFT and the $m=4$ tricritical Ising CFT. Second, we use these results to clarify the RG flow structure of the multicritical Ising sequence that formed the basis of the epsilon expansion analysis in Ref.~\cite{PatilLudwig2024}. For the tricritical Ising ground state under energy measurements, our results with logarithmic measurement-averaged entanglement support the RG flow at any weak measurement strength to 
a weak-measurement
fixed point.
For the critical Ising ground state under spin measurements, we instead find a direct RG flow toward the projective-measurement fixed point, 
which implies area-law entanglement at any finite measurement strength in the thermodynamic limit. 

Across both cases, the numerically extracted universal quantities and scaling exponents are in qualitative agreement with the epsilon expansion 
predictions obtained to one- and two-loop order where such predictions are available.

The rich criticality of these post-measurement ensembles is characterized through
a number of universal structures. The 
logarithmic part of the measurement-averaged von Neumann entanglement entropy defines an entanglement
effective central charge $c_{\rm eff}$. In the Euclidean spacetime description, the measurement layer defines a defect, and folding across this defect converts the problem into that of a boundary RG flow. 
At the resulting infrared boundary fixed point,
the effective Affleck-Ludwig boundary entropy $s_{\rm eff}$~\cite{PhysRevLett.67.161,PatilLudwig2024} is a universal quantity that is computed from the system-size independent contribution in the Shannon entropy
of the measurement record~\cite{PatilLudwig2024,PatilLudwig20251}.
Finally, we find signatures of multifractal scaling in the Born-averaged moments of correlation functions,
meaning that different moments are governed by independent scaling exponents rather than by integer multiples of a single scaling dimension, and whose relationship reflects the fact that they arise from a scaling form of an underlying probability distribution~\cite{PhysRevLett.128.050602,PatilLudwig2024,DuplantierLudwig1991,LUDWIG1990639}.
The multifractal convexity obeyed by these exponents is a hallmark of the non-unitarity arising from the randomness of measurement outcomes and is opposite to that in unitary field theories. These quantities together provide a detailed characterization of the measurement-dominated infrared behavior and of the crossover away from the clean multicritical Ising
points. 

The remainder of the paper is organized as follows. In Sec.~II, we introduce the O'Brien--Fendley chain and the weak-measurement protocols used for the critical Ising and tricritical Ising ground states. In Sec.~III, we review the replicated field-theory formulation of the setups, discuss the RG flow, and derive the crossover
scales
relating the microscopic measurement strength to the effective field-theory coupling. In Sec.~IV, we present the finite-size crossover analysis for the tricritical Ising ground state under energy measurements. In Sec.~V, we perform the corresponding analysis for the critical Ising ground state under spin measurements and show that the flow proceeds toward the projective-measurement fixed point. We conclude in Sec.~VI with a discussion of the resulting RG-flow picture for the multicritical Ising sequence subjected to measurements with a certain operator.

\par

\section{Model Setup}

We introduce the microscopic lattice model used to study post-measurement ensembles of critical ground states. The
measurements 
generate intrinsic Born-rule randomness in the resulting ensemble of wavefunctions and, in the cases we consider,
trigger a relevant RG flow. Our goal is to 
understand the universal properties of this post-measurement ensemble and characterize the RG flow away from the unmeasured fixed  point, and its crossover to a novel measurement-dominated infrared fixed point.   

\subsection{Hamiltonian}

We use the one-dimensional O'Brien--Fendley chain~\cite{PhysRevLett.120.206403} as a microscopic model that realizes both critical Ising and tricritical Ising 
 ground states within the same zero-temperature phase diagram. These two points correspond to distinct criticalities of Ising-type systems and provide natural lattice representatives of the $m=3$ and $m=4$ unitary minimal model CFTs. They therefore offer a controlled setting to compare how the two unmeasured quantum critical and tricritical Ising ground states
respond to the measurement perturbations considered below. The Hamiltonian is
\begin{equation}
\begin{aligned}
H &= 2H_{I} +\gamma H_3,  \\
H_{I} &= -\sum_{i=1}^{L} \left(\sigma^{z}_{i}\sigma^{z}_{i+1} + \sigma^{x}_{i}\right), \\
H_3 &= \sum_{i=1}^{L} \left(\sigma_i^x \sigma_{i+1}^z \sigma_{i+2}^z + \sigma_i^z \sigma_{i+1}^z \sigma_{i+2}^x\right),
\end{aligned}
\label{eq:obrien}
\end{equation}
where $\sigma_i^\alpha$ are Pauli matrices acting on spin-$1/2$ degrees of freedom on a chain of length $L$. We impose periodic boundary conditions, with sites $L+1$, $L+2$ in the sum 
understood modulo $L$. 
This unmeasured system
is disorder-free [i.e., it contains no (impurity-type) quenched disorder],
and is invariant under Kramers--Wannier duality.

The first term, $H_I$, is the one-dimensional critical quantum Ising Hamiltonian. The second term, $H_3$, is a three-spin interaction controlled by the coupling $\gamma>0$. Starting from the critical Ising point at $\gamma=0$, increasing $\gamma$ drives the model to a tricritical Ising point at $\gamma=\gamma_{tc}\approx0.856$. Beyond this point, the model enters a gapped phase. The tricritical point separates the second-order critical Ising transition line from the first-order transition, as shown schematically in Fig.~\ref{fig:model_setup}(a). Its universal 
behavior is known at the two critical points considered in this work: $\gamma=0$, described by the critical Ising CFT, and $\gamma=\gamma_{tc}$, described by the tricritical Ising CFT.

These two critical points serve as unmeasured (clean) reference states for the post-measurement ensembles studied in this work. In the absence of measurements, their universal properties are fixed by the corresponding unitary minimal model CFTs. The critical Ising ground state realizes the $m=3$ minimal model with central charge $c=1/2$, while the tricritical Ising ground state realizes the $m=4$ minimal model with central charge $c=7/10$. 
Correlation functions of local operators are governed by power-law decay: the scaling dimension associated with the spin operator in the critical Ising ground state is ${X}^{(\sigma)}=1/8$, while the scaling dimension associated with the energy operator~\cite{Nienhuis1982ExactTricrit,BELAVIN1984333,PhysRevLett.52.1575,DotsenkoFateev}~\footnote{ The properties of the corresponding microscopic operator in the O'Brien-Fendley model were established in~\cite{PhysRevB.101.045132}}
in the tricritical Ising ground state is ${X}^{(\mathcal E)}=1/5$. 

Introducing measurements converts each unmeasured critical ground state into an ensemble of post-measurement states, with outcomes sampled according to the Born rule. This ensemble carries intrinsic  measurement-induced randomness~\footnote{Whereas the underlying unmeasured Hamiltonian is of course free of any kind of randomness/disorder.}.
In the following subsection, 
we define two different weak-measurement protocols 
for the above two respective critical ground states, 
which will generate these post-measurement ensembles.
The comparison between the critical Ising and tricritical Ising points then allows us to probe how the post-measurement infrared behavior depends, respectively, on the underlying Ising and tricritical Ising criticality.

\subsection{Measurement protocols}

The action of a measurement layer on the critical ground state $|\psi_{\rm GS}\rangle$, shown in Fig.~\ref{fig:model_setup}(b), produces an (normalized) outcome-conditioned post-measurement state $|\psi_{\rm GS}(\vec{m})\rangle$. Here $\vec{m}$ denotes the measurement record, with outcomes sampled according to the Born probability $p_{\vec{m}}$. Each measurement record labels a  ``trajectory''
in the ensemble of post-measurement wavefunctions. Although, as stressed before, the underlying Hamiltonian is disorder-free, the Born-rule sampling introduces intrinsic randomness into this ensemble. As shown in Ref.~\cite{PatilLudwig2024},
measurements
can be relevant in the RG sense and can generate new  critical behavior that is characterized by their intrinsic randomness, which is clearly absent
in the unmeasured ground state. We study this  critical  behavior using two measurement protocols, one for the tricritical Ising ground state and one for the critical Ising ground state.

For the {\it tricritical} Ising ground state, we measure the local energy operator on alternating bonds. On the bond connecting sites $i$ and $i+1$, we define
\begin{equation}
E_{i,i+1}
=
\frac{1}{\sqrt{2}}
\left(
\sigma_i^z\sigma_{i+1}^z-\sigma_i^x
\right).
\label{eq:energy_op_tci}
\end{equation}
This operator is a lattice representative of the local energy field $\mathcal{E}$ of the tricritical Ising CFT. The operator $E_{i,i+1}$ is odd under the Kramers--Wannier transformation and for notational simplicity, we label the bond by its left site and write $E_i\equiv E_{i,i+1}$.

The two-site energy operator is Hermitian, $E_i^\dagger=E_i$, and involutory, $E_i^2=1$. It therefore has eigenvalues $\pm1$, each with a  two-fold degenerate~\cite{PatilLudwig2024} eigenspace. We can define the weak-measurement Kraus operators by
\begin{equation}
K_{i}(m_i)
=
\frac{1+m_i\lambda E_i}
{\sqrt{2(1+\lambda^2)}} ,
\label{eq:kraus_operator_tci}
\end{equation}
where $m_i=\pm1$ is the measurement outcome and $\lambda\in[0,1]$ is the measurement strength. The limits $\lambda=0$ and $\lambda=1$ correspond respectively to no measurement and to a projective measurement of $E_i$. Using $E_i^\dagger=E_i$ and $E_i^2=1$, these operators satisfy the POVM completeness condition
\begin{equation}
\sum_{m_i=\pm1}
K_i^\dagger(m_i)K_i(m_i)
=
1 .
\label{eq:povm}
\end{equation}
For a single measured bond, the Born probability of outcome $m_i$ is
\begin{equation}
p_{m_i}
=
\langle\psi_{\rm GS}|
K_i^\dagger(m_i)K_i(m_i)
|\psi_{\rm GS}\rangle ,
\label{wq:born_probability}
\end{equation}
and the normalized post-measurement state is
\begin{equation}
|\psi_{\rm GS}(m_i)\rangle
=
\frac{K_i(m_i)|\psi_{\rm GS}\rangle}
{\sqrt{p_{m_i}}}.
\label{eq:kraus_action}
\end{equation}

For a chain of even length $L$, we measure $L/2$ alternating bonds. We choose the measured set to be alternating links ${1,3,\ldots,L-1}$. Since these bond operators have disjoint support, the corresponding Kraus operators commute. Thus, the final state for a fixed measurement record is independent of the order in which the measurements are applied. For a trajectory $\vec{m}$, the full Kraus operator is
\begin{equation}
K(\vec{m})
=
\prod_{i\in \rm{odd}} K_i(m_i).
\label{eq:kraus_vec}
\end{equation}
and the Born probability of the full measurement record is
\begin{equation}
p_{\vec{m}}
=
\langle\psi_{\rm GS}|
K^\dagger(\vec{m})K(\vec{m})
|\psi_{\rm GS}\rangle ,
\end{equation}
with the normalized post-measurement state given by
\begin{equation}
|\psi_{\rm GS}(\vec{m})\rangle
=
\frac{K(\vec{m})|\psi_{\rm GS}\rangle}
{\sqrt{p_{\vec{m}}}}.
\label{eq:kraus_full_action}
\end{equation}

Similarly, for the critical Ising ground state, we use a spin-measurement protocol in which $\sigma_i^z$ is measured on every site. Since $(\sigma_i^z)^\dagger=\sigma_i^z$ and $(\sigma_i^z)^2=1$, we define
\begin{equation}
K_{i}(m_i)
=
\frac{1+m_i\lambda\sigma_i^z}
{\sqrt{2(1+\lambda^2)}} ,
\label{eq:kraus_operator_ising}
\end{equation}
with $m_i=\pm1$. These Kraus operators satisfy the same POVM completeness condition. For a measurement record $\vec{m}$, the full Kraus operator is $K(\vec{m})
=
\prod_{i=1}^{L}K_i(m_i)$,
with Born probability $p_{\vec{m}}
=
\langle\psi_{\rm GS}|
K^\dagger(\vec{m})K(\vec{m})
|\psi_{\rm GS}\rangle$,
and normalized post-measurement state given by Eq.~\ref{eq:kraus_full_action}.

We note that for the respective critical states one can consider
other 
measurement protocols, resulting in a universal
long-distance behavior of the post-measurement ensemble
that is governed by the same field theory as that for above mentioned measurement protocols.
For example, the long-distance physics in the case of weak measurements with the $\sigma_i^x$ operator on alternating sites of the tricritical Ising ground state
is expected to be governed by the same replica field theory as that for the protocol in Eq.~\eqref{eq:kraus_operator_tci}; or the field theory description of the protocol where we perform weak measurements with the $\sigma_i^z$ operator on alternating sites of the critical Ising ground state is expected to be the same as that for the above protocol where we perform weak measurements with $\sigma_i^z$ on all sites of the Ising critical ground state.

In the remainder of this work, we focus on energy measurements with the Kraus operators in Eq.~\eqref{eq:kraus_operator_tci} on alternating bonds of the tricritical Ising ground state and $\sigma^z$ measurements on all sites for the critical Ising ground state.

\begin{figure}[t]
  \centering
  \includegraphics[width=0.35\textwidth]{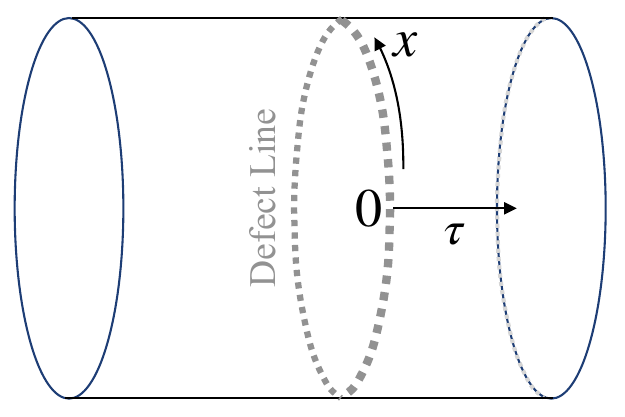}
  \caption{\textbf{Measurement defect line.} The action of performing measurements on critical ground state with periodic boundary is equivalent to introducing random defect on the zero time slice of the unmeasured field theory on a cylindrical space-(imaginary) time geometry (See e.g. \cite{PhysRevX.13.021026,LeeJianXu,PatilLudwig2024}). The probability of the randomness is proportional to the partition function of the defect theory itself, which follows from Born's rule.}

    \label{fig:measurement_defect}
\end{figure}

\section{Replica field theory and RG flow\label{SecReplicaTheoryandRGFlow}}

The ground states of critical
 quantum Ising and tricritical Ising Hamiltonians
subjected respectively to spin and energy measurements, can be studied within a 
replicated theory for replica copies of
 the generalized multicritical Ising 
Landau-Ginzburg fixed-point action~\cite{osti_6647980} perturbed by a certain replica interaction~\cite{PatilLudwig2024}. 
The purpose of using the {\it generalized} Landau-Ginzburg fixed-point action 
is  to introduce a small parameter $\epsilon$ to be used in the epsilon expansion.
The unmeasured critical Ising and tricritical Ising ground states are part of the series of $(1+1)D$ Ising multicritical points described by the following Landau-Ginzburg-Zamolodchikov action~\cite{osti_6647980}
\begin{equation}
    S_{*} = \int d\tau \int dx \left(\frac{1}{2}(\partial_{\tau}\phi)^2 + \frac{1}{2}(\partial_{x}\phi)^2 + g^{*}_{m-1}\phi^{2(m-1)} \right),
\label{eq:multicritical_ising_action}    
\end{equation}
where $\phi(x, \tau)$ is a real scalar field, $m=3$ and $m=4$ correspond to the critical Ising {(``$\phi^4$ theory'')} and the tricritical Ising  {(``$\phi^6$ theory'')} ground state, respectively~\footnote{In the present case of two dimensions there is no meaning to a perturbative study of the Landau-Ginzburg action about the Gaussian theory, since the field $\phi$ is dimensionless by naive 
power-counting, and a non-perturbative tool is needed. This is provided in Ref.~\onlinecite{osti_6647980}
where it is shown that non-perturbative field identifications following from the exact equations of motion are exactly those obtained from the corresponding $A$-series unitary minimal model CFT.}.
As illustrated in Fig.~\ref{fig:measurement_defect}, in the field theoretic language, measurements on a ground state of length $L$ introduce a random defect perturbation localized on the zero-time slice in the corresponding critical CFT action defined on the space-(imaginary)-time cylindrical geometry of circumference $L$.
By using the replica trick and averaging over the measurement outcomes, the 
generalized replica action derived in Ref.~\cite{PatilLudwig2024} is given by
\begin{equation}
\begin{aligned}
    -\mathsf{S} &= -\sum_{a =1}^{R}S_{*}^{a} + \Delta \int_{-\infty}^{\infty} dx~\sum_{\substack{a,b=1 \\a\neq b}}^{R}\chi^{(a)}(x,0)\chi^{(b)}(x,0),     
\end{aligned}
\label{replica_action_full}
\end{equation}
where $S_*$ is
the Landau-Ginzburg-Zamolodchikov action 
for 
the unperturbed $(m-2)^{\text{th}}$
$\mathbb{Z}_2$ Ising multicritical point [Eq.~\eqref{eq:multicritical_ising_action}], the
coupling $\Delta \in [0,\infty)$ [Eq.~\eqref{replica_action_full}] denotes the effective measurement strength, $R$ is the number of replicas, and (the colon describes normal ordering)
\begin{equation}
    \chi = :\phi^{m-2}:  \label{EqChi}
\end{equation}
is the generalized local measurement field.
For $m=4$, the action $S_*$ corresponds to the tricritical Ising CFT and the field $\chi=:\phi^2:$ represents the energy field $\mathcal{E}$ of the CFT, and the replica action in Eq.~\eqref{replica_action_full} in the replica $R\rightarrow1$ limit governs the tricritical Ising 
critical ground state subjected to energy measurements. 
On the other hand, for
$m=3$, the action $S_*$ corresponds to the critical Ising CFT and the
field $\chi=\phi$ represents the spin field $\mathcal{S}$ of the CFT, and the replica action in Eq.~\eqref{replica_action_full} in the replica $R\rightarrow1$ limit governs the Ising critical ground state subjected to spin measurements. The 
scaling dimension 
$X_{\chi}$ of the field
$\chi  = 
:\phi^{m-2}:$
in any $(m-2)^{\text{th}}$ multicritical Ising CFT ($m\geq 3$), also known as  the $m^{\text{th}}$ ($A$-series) Virasoro minimal model CFT,
is 
known exactly from $2D$ CFT literature~\cite{BELAVIN1984333,PhysRevLett.52.1575,DiFrancesco:1997nk}~\footnote{The field $\chi$ corresponds to the Kac's table primary field $\varphi_{1,2}$ of the $m^{\text{th}}$ minimal model ($A$-series) Virasoro CFT governed by action in Eq.~\eqref{eq:multicritical_ising_action}}
\begin{equation}
    X_{\chi} = \frac{1}{2} - \frac{3}{2(m+1)},
\label{eq:chi_sclalingdimension}    
\end{equation}
which reduces to the spin and energy scaling dimension $X^{(\sigma)} = 1/8$ and $X^{(\mathcal{E})} = 1/5$ of $m=3$ critical Ising and $m=4$ tricritical Ising 
model CFTs, respectively. The coupling $\Delta$ is relevant in the RG sense for any $m\geq 3$ and the corresponding RG eigenvalue is given by
\begin{equation}
    y_{\Delta} = 1 -2X_{\chi} = \frac{3}{m+1},
\label{eq:rg_eigenvalue}
\end{equation}
that stays positive for all (integer) $m\geq 3$.
Thus 
measurements of $\chi$ [Eq.~\eqref{EqChi}] introduce a relevant RG
flow at all finite 
$m$ in Fig.~\ref{fig:model_setup}(c). 
This as well defines a small parameter $\epsilon = 3/(m+1)$ to carry out a controlled RG expansion about 
$m = \infty$
where the perturbation becomes marginally irrelevant~\cite{PatilLudwig2024,JengLudwig}. The 
perturbative RG $\epsilon$-expansion has previously revealed a measurement-dominated infrared fixed point at finite measurement strength $\Delta_{*}$, shown in Ref.~\cite{PatilLudwig2024}, that shows a rich and complex critical behavior which differs
drastically
from their unmeasured CFT. We numerically investigate the presence of such a fixed point in the 
ground states of the tricritical and critical quantum Ising models subjected to weak measurement.

Before presenting the numerical results, we first obtain the RG crossover scale going to be used in our
systematic finite-size scaling analysis. To this end, we briefly review the relation between the effective measurement strength $\Delta$ appearing in the 
replica field theory and the measurement strength $\lambda$ used in the lattice Kraus operators. 
Note that the sum  over the discrete measurement outcomes $m_i= \pm 1$
of replicated product of Kraus operator $(K^{\dagger}_i (m_i)K_i (m_i))^{\otimes R}$ was shown in~\cite{PatilLudwig2024}  to be given by
 \begin{equation}
\sum_{m_{i}=\pm 1} (K^{\dagger}_i (m_i)K_i (m_i))^{\otimes R}= \exp{\{\tilde{\lambda}^2\sum_{\substack{a,b=1,\\a\neq b}}^R\chi_i^{(a)}\chi_i^{(b)}+\mathcal{O}(\tilde{\lambda}^4)\}}.
 \end{equation}
 where $\tilde{\lambda}=\tanh^{-1}{(\lambda)}$ and $\chi_i=\sigma_i^{z}$ or $E_i$ is the local measurement operator. The 
 factor
 $(\tanh^{-1}{(\lambda)})^2$ couples to the term with lowest number of replicated fields $\chi_i^{(a)}$,
 and is hence the most RG relevant perturbation to the (replicated) critical theory. In particular, when we move to the continuum language, the factor
 $(\tanh^{-1}{(\lambda)})^2$ is proportional to the field theory coupling constant $\Delta$ in Eq.~\eqref{replica_action_full}.

Since we know that coupling $\Delta$ is relevant under RG with the RG eigenvalue given in Eq.~\ref{eq:rg_eigenvalue}, for a finite system size $L$, we  obtain an RG crossover scale  $\xi$ from the dimensionless ratio
\begin{equation}
    L\Delta^{1/y_{\Delta}} \propto L(\tanh^{-1}(\lambda))^{2/y_{\Delta}}, 
\label{eq:rg_crossover}    
\end{equation}
 with $y_{\Delta} = 3/4$ and $3/5$ for ground state of $m=3$ critical Ising and $m=4$ tricritical Ising minimal model CFTs, respectively. We can define  $L(\tanh^{-1}(\lambda))^{2/y_{\Delta}} = L/\xi$, where $\xi = (\tanh^{-1}(\lambda))^{-2/y_{\Delta}}$ is the crossover length scale and $L \gg \xi$ defines the infrared (IR) region and $L \ll \xi$ denotes the ultraviolet (UV) regime. In the sections below, we use the so-defined RG crossover scale $\xi$ to investigate the measurement-dominated criticality in the infrared regime of tricritical Ising and critical Ising  ground states  subjected to their respective weak measurement protocol and establish the RG flow shown in Fig.~\ref{fig:model_setup}(c).

\begin{figure*}[t!]
	\centering
	\includegraphics[width=\textwidth]{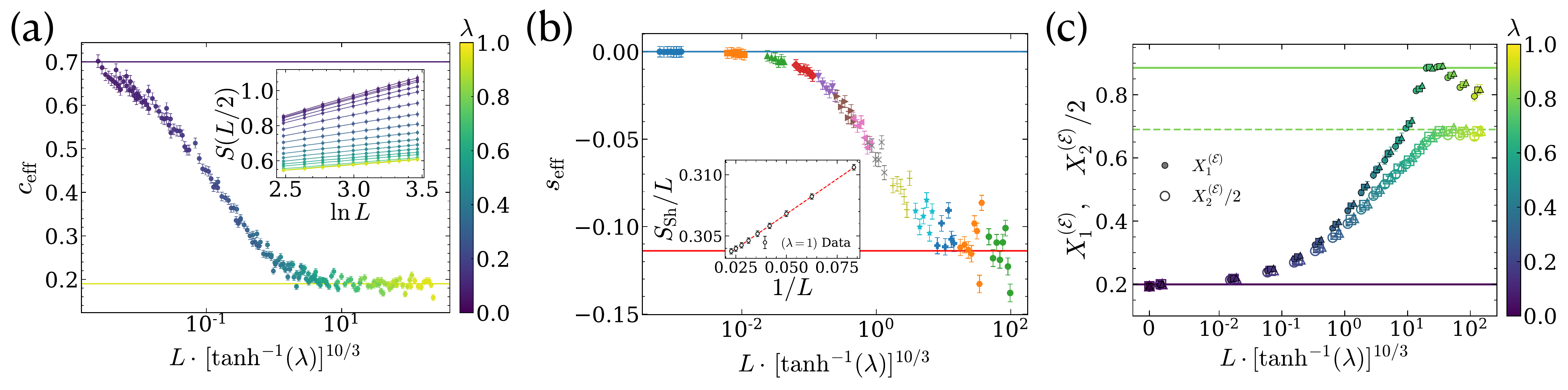}
	\caption{\textbf{Universality of tricritical Ising subjected to weak energy measurement.}~We show finite-size RG crossover collapses of universal quantities and scaling exponents computed from the post-measurement TCI ground state with periodic boundary conditions. The data collapse as a function of $L/\xi$, where $\xi=(\tanh^{-1}\lambda)^{-10/3}$ is the crossover length associated with the RG relevant perturbation induced by measurements. The saturation of universal quantities and scaling exponents in the regime $L/\xi\gg1$ provides evidence for a measurement-dominated infrared fixed point. (a) Entanglement effective central charge $c_{\rm eff}$ extracted from the half-system von Neumann entanglement entropy. The inset shows the logarithmic scaling with system size $L$. The collapse shows a flow from the unmeasured TCI value $c=7/10$ to a saturated value $c_{\rm eff}\simeq0.19$ near the infrared fixed point. (b) Effective Affleck-Ludwig boundary entropy $s_{\rm eff}$ extracted from the finite-size scaling of the Shannon entropy density $S_{\rm Sh}/L$ of the measurement record. The data suggest a monotonically decreasing flow from $s_{\rm eff}=0$ to $s_{\rm eff}\simeq -0.114$ in the projective limit, obtained from the finite-size fit shown in the inset, $S_{\mathrm{Sh}}/L=-(-0.114)(1/L)+0.301$. Each data point is averaged over $8\times10^7$ samples. (c) Multifractal scaling of the Born-averaged connected energy-energy correlation function. The first- and second-moment exponents, $X_{1}^{(\mathcal{E})}$ and $X_{2}^{(\mathcal{E})}$, flow from their unmeasured TCI values, $1/5$ and $2/5$, to the infrared  values $0.885(9)$ and $1.38(5)$, respectively. The saturation of both exponents at large $L/\xi$ supports the emergence of the measurement-dominated infrared fixed point. The TCI ground state for each system size was obtained using DMRG with a truncation cutoff  of $10^{-10}$. In all panels, reference lines mark the unmeasured TCI values and the saturated infrared values.
    }	\label{fig:universality_tci}
\end{figure*}

\section{Tricritical Ising Ground State Under Energy Measurements}

We numerically investigate the 
critical behavior that emerges in the 
post-measurement ensemble of the tricritical Ising  (TCI)  ground state subjected to weak energy measurements. The tricritical Ising  model  represents the $m=4$ unitary minimal model CFT with central charge $c=7/10$. The field theory description for the problem of the tricritical Ising ground state subjected to weak energy measurements corresponds to the case of $m=4$ in the generalized replica field theory
of Eqs.~\eqref{eq:multicritical_ising_action} and \eqref{replica_action_full}
in the $R\rightarrow1$ replica limit discussed in the previous section.
 Using the epsilon expansion of
Ref.~\cite{PatilLudwig2024}
 where $\epsilon=3/(m+1)$],
the infrared behavior for the post-measurement ensemble is expected to be governed by a measurement-dominated fixed point, 
which gets reflected in several universal quantities and scaling exponents. We focus on such universal quantities by systematic numerical analysis of the measurement-averaged bipartite entanglement entropy, which gives access to the entanglement effective central charge; the Shannon entropy of measurement record, which contains 
the
universal effective Affleck-Ludwig boundary entropy; and  moments of  Born-averaged two-point correlation functions, which reveal signatures of multifractality in the post-measurement ensemble. We use finite-size RG crossover collapses with the scaling variable $L(\tanh^{-1}\lambda)^{2/y_{\Delta}}=L/\xi$, where $y_{\Delta}=3/5$ corresponds
to the $m=4$ TCI case
in Eq.~\ref{eq:rg_eigenvalue}. The resulting collapse shows that these observables consistently approach universal saturated values in the infrared regime, providing evidence for a measurement-dominated fixed point in the weakly measured TCI ground state.

\subsection{Effective Central Charge ($c_{\rm eff}$)}

We compute the measurement-averaged bipartite von Neumann entanglement entropy between the two halves of the system for the post-measurement TCI ground state with
 periodic boundary conditions. 
We find that the
entanglement grows logarithmically with system size $L$ for all measurement  strengths $0\leq\lambda\leq 1$, where $\lambda =0$  corresponds to no measurement and $\lambda =1$ 
 corresponds to projective measurement on alternating links.
Note that even in the projective limit (on alternating links), this
two-site local energy measurement does 
not collapse the state into a trivial 
product
state and the residual entanglement turns out to be logarithmic. The logarithmic measurement-averaged von Neumann entanglement entropy is characterized by the universal entanglement
effective central charge $c_{\rm eff}$ defined by
\begin{equation}
    \overline{S(L_{A})} = \frac{c_{\rm eff}}{3} \log (L_{A}) + \mathcal{O}(1).
\label{eq:entanglement_formula}    
\end{equation}

We set $L_A = L/2$ and extract $c_{\rm eff}$~\footnote{We note that the \textit{entanglement} effective central charge $c_{\text{eff}}$ here should be distinguished from the \textit{Casimir} effective central charge that characterizes universal information content in Shannon entropy of measurement record in $(1+1)D$ monitored deep quantum circuits or monitored $2D$ Rokhsar-Kivelson wavefunctions. (See 
Ref.~\cite{PhysRevLett.128.050602},
and
{comments on this distinction},
e.g., 
in
Refs.~\cite{PhysRevB.109.014303,puetz2025flownishimori,wang2025decoherenceselfdual,PatilLudwig20251}.)} from the local slope of the measurement-averaged entanglement entropy as a function of $\log L$ (see inset of Fig.~\ref{fig:universality_tci}(a)) and the resulting $c_{\rm eff} (L,\lambda)$ values show a good RG crossover collapse as a function of the TCI scaling variable $L/\xi$, as shown in Fig.~\ref{fig:universality_tci}(a). In the UV regime, $L/\xi\ll1$, the data approach the unmeasured TCI value $c=7/10$ whereas in the infrared regime the flow saturates to $c_{\rm eff}=0.19(4)$, indicating a measurement-dominated fixed point with nonzero logarithmic entanglement. Note that
the values predicted by the one-loop epsilon expansion result for the entanglement effective central charge in Ref.~\cite{PatilLudwig2024} are negative at $m=4$, 
and a  higher loop order computation would be required (but has not been performed to date) 
to obtain a sensible value.

To further test this behavior at larger system sizes, we also perform finite-size entanglement scaling with {\it open boundary conditions}, where DMRG allows access to larger
system sizes $L$. As shown in Fig.~\ref{fig:entanglement_scaling}(b), the measurement-averaged entanglement entropy remains logarithmic for all measurement strengths $\lambda$, with no crossover to an area-law state. Moreover, we expect the same behavior
to hold in the post-measurement ensemble obtained by performing weak (non-projective) measurements with the $\sigma^x$ operator on each site of the TCI ground state, 
as the local
energy field $\mathcal{E}$ is also the leading continuum field representing the lattice operator $\sigma^x_i$.
We have checked this expectation for certain system sizes.


\subsection{Effective Affleck-Ludwig Boundary Entropy ($s_{\rm eff}$)}

The measurement layer defines a defect line in the unmeasured two-dimensional conformal
field theory on the cylindrical space--imaginary-time geometry, as illustrated in Fig.~\ref{fig:measurement_defect}. Upon folding across the defect, the measurement layer becomes a boundary condition  on two copies of the unmeasured bulk CFT. At the measurement-dominated fixed point, this boundary is expected to be conformally invariant and contributes a universal constant term to the Shannon entropy of the measurement record, $S_{\rm Sh}$, which can be  interpreted~\cite{PatilLudwig2024} as the measurement-averaged free energy of this defect, and its universal constant piece defines the effective Affleck-Ludwig boundary entropy $s_{\rm eff}$.

The Shannon entropy takes the following finite-size~\cite{PatilLudwig2024} form 
in the scaling regime
\begin{equation}
S_{\rm Sh}
=
-\sum_{\vec{m}} p_{\vec{m}}\ln p_{\vec{m}}
=
f_0 L - s_{\rm eff},
\label{eq:boundary_entropy}
\end{equation}
where $f_0$ is a nonuniversal extensive contribution and $s_{\rm eff}$ is the universal boundary contribution. In the unmeasured limit, $s_{\rm eff}=0$ for the trivial reason that no nontrivial measurement defect is present. Equivalently, from Eq.~\ref{eq:boundary_entropy}, the Shannon entropy density scales as $\frac{S_{\rm Sh}}{L}=f_0 - \frac{s_{\rm eff}}{L}$.

We extract $s_{\rm eff}$ from the finite-size slope of $S_{\rm Sh}/L$ as a function of $1/L$, shown in the inset of Fig.~\ref{fig:universality_tci}(b) with projective limit data. The resulting $s_{\rm eff}(L,\lambda)$ values show a finite-size RG crossover collapse with $L/\xi$ and decrease monotonically from the trivial unmeasured value $s_{\rm eff}=0$ to $s_{\rm eff}=-0.114$ in the infrared regime. This decrease is consistent with the
$g$-effective theorem
of Ref.~\cite{PatilLudwig20251}. On comparison, the saturated numerical value is also close to the perturbative RG prediction $s_{\rm eff}=-0.0888$ obtained to order $\epsilon^3$  in the epsilon expansion, as mentioned in Table.~\ref{tab:table_tci}.

\subsection{Multifractality}

Within the 
epsilon expansion of Ref.~\cite{PatilLudwig2024}, the measurement-dominated fixed point 
 exhibits multifractal scaling, where different moments
of a measurement-averaged correlation function decay with independent scaling exponents, rather than  with integer multiples of a single scaling dimension. This produces a hierarchy of  independent  exponents in the correlation function distribution.
 More precisely, exponents $X_N$ characterizing the power law decay of the $N$-th moment in a multifractal theory must satisfy a convexity condition $X_{N+M} \leq X_N + X_M$, which is a consequence of the fact that the entire hierarchy of exponents $X_N$ arises from an underlying universal scaling function for the probability distribution. Hierarchies of critical exponents in unitary field theories such as e.g. the 2D or 3D Ising model~\footnote{ such as (normal ordered) powers $:\phi^N:$ of the Landau-Ginzburg field $\phi$}, satisfy~\cite{DuplantierLudwig1991} opposite convexity conditions  $X_{N+M} \geq X_N + X_M$. The convexity of the exponents we observe is a hallmark of the underlying non-unitarity arising from the randomness of measurement outcomes: Below, we numerically establish $X^{(\mathcal{E})}_2 < 2 X^{(\mathcal{E})}_1$, which thus reflects the multifractal nature of the measurement-dominated fixed point we observe.

 Specifically, we test this prediction using the absolute value of the measurement-averaged connected energy correlation in the post-measurement ensemble. The absolute value is taken to avoid cancellations from sign fluctuations between different measurement outcomes. For a periodic system, the $N^{\rm th}$ moment is expected to scale with the chord distance as

\begin{equation}
    \overline{|\langle\delta E_i \delta E_j\rangle_{\vec{m}}|^{N}} \propto \frac{1}{\left(\frac{L}{\pi}\cdot\sin\left(\frac{\pi(|r-r_0|)}{L}\right)\right)^{2X_{N}^{(\mathcal{E})} }},
\label{eq:tci_correlation}    
\end{equation}
where $\delta E_i=E_i-\langle E_i\rangle_{\vec{m}}$ and the overline denotes averaging over measurement outcomes with Born probabilities $p_{\vec{m}}$. The lattice sites $i$ and $j$ correspond to position $r$ and $r_0$, respectively. We extract the power law exponent $X^{\mathcal{(E)}}_{N}$ from the log-log linear fit for the first and second moment of the measurement-averaged connected energy correlation function in the post-measurement ensemble. Since we are working with periodic boundary conditions, we demonstrate such power law fits by plotting $\ln(\overline{|\langle \delta E_i \delta E_j\rangle_{\vec{m}}|^N})$ against $\ln\left(\frac{L}{\pi}\cdot \sin\left(\frac{\pi}{L} (|r -r_0|)\right)\right)$ for $N=1,2$ in the appendix Fig.~\ref{fig:tci_multifractality}. 

Near the projective limit, the connected energy correlator develops an odd-even staggered structure in each measurement trajectory due to the microscopic pattern of measured and unmeasured links. This staggered component is a lattice-scale effect of the measurement protocol, rather than a feature of the universal long-distance correlations. It arises because the same local energy operator is first measured on a subset of links and then used to define the connected two-point function in the post-measurement state. As a result, correlations involving directly measured links are strongly suppressed relative to those involving unmeasured links, producing an alternating structure in the correlator in every measurement trajectory $\vec{m}$.
At the projective limit, the measured links, say $j$, have vanishing correlation $\langle E_i E_j\rangle_{\vec{m}}-\langle E_i\rangle_{\vec{m}}\langle E_j\rangle_{\vec{m}}
= \langle E_i\rangle_{\vec{m}}-\langle E_i\rangle_{\vec{m}} =0$. However, links that are not directly measured retain nontrivial long-range
correlations with each other. This is consistent with the universal behavior extracted from both $c_{\rm eff}$, characterizing logarithmic entanglement growth, and also the effective boundary entropy $s_{\rm eff}$ at $\lambda=1$. 
 
In our numerical analysis, we fix an unmeasured reference link at $i=L/2$ and sweep over the remaining links to compute the connected correlator with respect to this reference link. To smooth the odd-even lattice modulation near the projective limit, we average neighboring odd and even links of the Born-averaged correlator, i.e. we average the values at $j$ and $j+1$ in $\overline{|\langle\delta E_{L/2}\delta E_j\rangle_{\vec{m}}|^N}$. This removes the leading microscopic staggered component while preserving the long-distance scaling used to extract the universal exponents.    

The exponents $X_{N}^{(\mathcal E)}$ obtained from the first and second moments obey the RG crossover collapse as a function of $L/\xi$, as shown in Fig.~\ref{fig:universality_tci}(c). In the UV regime, $L/\xi\ll1$, they approach the unmeasured TCI values $X_{1}^{(\mathcal E)}=\frac{1}{5}$ and $X_{2}^{(\mathcal E)}=\frac{2}{5}=2\times \frac{1}{5}$. At large $L/\xi$, the exponents saturate to distinct infrared values, $X_{1}^{(\mathcal E)}=0.885(9)$ and $X_{2}^{(\mathcal E)}=1.38(5)$, indicating multifractal scaling at the measurement-dominated fixed point. The first-moment exponent is more sensitive to the staggered lattice effect discussed above; nevertheless, its saturated value remains close to the two-loop epsilon expansion result reported in Table~\ref{tab:table_tci}.

\begin{table}[h!]
\caption{\label{tab:table_tci}Universal exponents obtained for the measurement-dominated infrared fixed point using numerical and controlled
perturbative RG epsilon expansion approaches for ground state of tricritical Ising subjected to weak energy measurement.}

\begin{tabular}{ 
|p{2.0cm}||p{1.1cm}|p{1.7cm}|p{1.7cm}|p{1.5cm}| }
 \hline
 \multicolumn{1}{|c||}{Tricritical Ising} & \multicolumn{4}{c||}{Exponents/Universal numbers} \\
 \hline
 \multicolumn{1}{|c||}{}& $c_{\rm{eff}}$ & $s_{\rm{eff}}$ & $X_{1}^{(\mathcal{E})}$ & $X_{2}^{(\mathcal{E})}$ \\
 \hline
 Numerics & 0.19(4) & -0.114(8) & 0.885(9) & 1.38(5) \\
 $\epsilon$-expansion
 & $< 0$ & -0.0888 {\tiny \hspace{0.7cm}(1-loop)} & 1.16 {\tiny \hspace{0.8cm}(2-loop)} & 1.24 {\tiny \hspace{0.7cm}(2-loop)} \\  
 \hline
\end{tabular}

\end{table}

\begin{figure*}[t!]
	\centering
    \includegraphics[width=\textwidth]{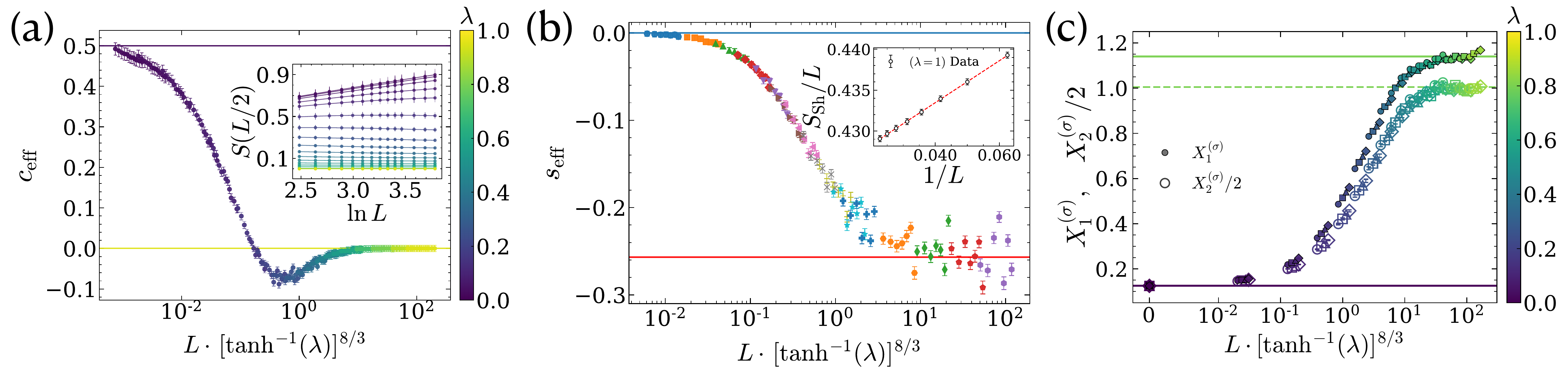}
    \caption{\textbf{Universality of critical Ising subjected to weak spin measurement.}
    We show finite-size RG crossover collapses of universal quantities and scaling exponents computed from the post-measurement Ising ground state with periodic boundary conditions. The data collapse as a function of $L/\xi$, where $\xi=(\tanh^{-1}\lambda)^{-8/3}$ is the crossover length associated with the relevant spin-measurement perturbation about the unmeasured critical Ising fixed point. The collapse indicates a direct flow toward the projective-measurement fixed point. Analogous to the TCI RG crossover, the universal quantities and scaling exponents show saturation near the infrared regime at large $L/\xi$.~(a) 
    Entanglement effective central charge $c_{\rm eff}$ extracted from the half-system von Neumann entanglement entropy. The inset shows the entanglement scaling with system size $L$: at weak measurement and finite size it retains the logarithmic form of the critical Ising state, while at strong measurement it crosses over to area-law behavior. The extracted $c_{\rm eff}$ flows from the unmeasured Ising value $c=1/2$ to $c_{\rm eff}=0$, marking the area-law phase at the projective fixed point.~(b) Effective 
    Affleck-Ludwig boundary entropy $s_{\rm eff}$ extracted from the finite-size scaling of the Shannon entropy density $S_{\rm Sh}/L$ of the measurement record. The data show a monotonic decrease from $s_{\rm eff}=0$ in the unmeasured limit to $s_{\rm eff}\simeq -0.257$ in the projective limit, obtained from the finite-size fit shown in the inset, $S_{\mathrm{Sh}}/L=-(-0.257)(1/L)+0.423$. Each data point is averaged over $8\times10^7$ samples.~(c) Multifractal scaling of the Born-averaged connected spin-spin correlation function. The first- and second-moment exponents, $X_{1}^{(\sigma)}$ and $X_{2}^{(\sigma)}$, flow from their unmeasured Ising values, $1/8$ and $1/4$, to the infrared values $1.14(5)$ and $2.01(6)$, 
     respectively. The exponents saturate at large $L/\xi$ before reaching the projective fixed point. The critical Ising ground state for each system size was obtained using DMRG with a truncation cutoff  of $10^{-10}$.  In all panels, reference lines mark the unmeasured Ising values and the saturated infrared values.
}
	\label{fig:universality_ci}
\end{figure*}

\section{Ising Critical Ground State Under Spin Measurements}
We now discuss the case of the Ising critical ground state under spin measurements using a systematic finite-size RG crossover analysis that is analogous to the analysis performed above for the TCI ground state with energy measurements. The critical Ising point corresponds to the $m=3$ unitary minimal model CFT with central charge $c=1/2$. The perturbative
$\epsilon=\frac{3}{m+1}$ expansion analysis discussed in Sec.~\ref{SecReplicaTheoryandRGFlow} 
predicts, to any finite order in $\epsilon$, a 
weak
measurement 
fixed point characterized for example by logarithmic entanglement entropy for all integer $m\geq3$. However, our numerical results for the $m=3$ critical Ising ground state subjected to weak $\sigma^z$ measurements instead indicate a direct flow toward the projective-measurement fixed point, with area law entanglement entropy. On the other hand, note that for the case of $m=4$, i.e. the tricritical Ising ground state under energy measurements, our numerical results are consistent with the logarithmic entanglement predicted by the epsilon expansion with $\epsilon=\frac{3}{m+1}$. A natural possibility suggested by our numerics in this section is then that the Ising case ($m=3$) with spin measurements is the limiting case of the epsilon expansion where the 
weak-measurement 
fixed point predicted by the perturbative epsilon expansion collides with the 
projective-measurement 
fixed point, and they are one and the same fixed point. This is illustrated in Fig.~\ref{fig:model_setup}(c).

Heuristically, this direct RG flow to the projective limit can be understood by noting that the generalized replica theory discussed in Sec.~\ref{SecReplicaTheoryandRGFlow}, which forms the basis of the epsilon expansion in Ref.~\cite{PatilLudwig2024}, 
corresponds
for odd 
$m>3$
to a problem of performing measurements with a certain {\it subleading} spin operator on the ground state of the $m^{\text{th}}$ 
conformal minimal model CFT. 
However, in the case of $m=3$ ($\epsilon=3/4$), i.e. the Ising case, and only in this case, the corresponding measurement operator actually turns out to be the \textit{leading} spin operator from the unmeasured CFT ground state. 
Since, in the latter case, we are then
performing weak measurements 
with the order-parameter, the
most-relevant field in the Ising CFT, the observed flow directly to the projective limit is a 
reasonable RG flow.
 We also note that the RG-flow structure illustrated in Fig.~\ref{fig:model_setup}(c) for different $m \geq 3$ conformal minimal model CFTs suggests that even though any {\it finite} loop-order epsilon expansion could, by construction (unless resummed to all loop-orders), obviously never yield a vanishing entanglement effective central charge, reflecting the numerically observed structure of area-law entanglement in the Ising case, it is well conceivable that the expansion would capture other nontrivial universal properties~\footnote{unless a particular obstruction could be identified at some value  $3 < m_* <4$ that prevents the smooth continuation down to $m=3$. This possibility is at odds with the above heuristic argument suggesting a direct RG flow to the projective-limit only at $m=3$ and not at higher $m>3$ in the epsilon expansion. Further investigation of this possibility, and its relationship with the heuristic argument, will be deferred to future work.}. This includes, e.g., the effective Affleck-Ludwig boundary entropy of the critical Ising ground state under spin measurements, by evaluating the $\epsilon$ ($= 3/(m+1)$) expansion result at $m=3$.

For our systematic finite-size RG crossover analysis in this section, the collapse is organized by the scaling variable $L(\tanh^{-1}\lambda)^{8/3}$, obtained from the RG eigenvalue $y_\Delta=3/4$ for $m=3$, corresponding to critical Ising,
using Eq.~\ref{eq:rg_eigenvalue}.

\subsection{Effective Central Charge ($c_{\rm eff}$)}

The universal entanglement effective central charge $c_{\rm eff}$ is extracted from the bipartite von Neumann entanglement entropy $S(L_A)$ between two equal halves of the system, $L_A/L=1/2$, with periodic boundary conditions. For finite system size $L$, the entanglement entropy exhibits logarithmic scaling at weak measurement strength,
consistent with Eq.~\ref{eq:entanglement_formula}, followed by a crossover to area-law behavior above a finite-size crossover scale $\lambda_s$. This behavior is shown in Appendix Fig.~\ref{fig:entanglement_scaling}(a) and in the inset of Fig.~\ref{fig:universality_ci}(a). The crossover scale $\lambda_s$ drifts toward $\lambda=0$ as $L$ increases. This drift is consistent with the RG crossover collapse in Fig.~\ref{fig:universality_ci}(a), where the large-$L/\xi$ regime approaches $c_{\rm eff}=0$. These results indicate that, in the thermodynamic limit, any finite spin measurement strength $\lambda$ drives the critical Ising ground state toward an area-law phase.

The RG flow of $c_{\rm eff}$ starts from the unmeasured Ising value $c=1/2$ in the UV regime and decreases as the system crosses over away from the critical Ising fixed point. Near the crossover regime, where $L$ and $\xi$ are comparable, the finite-size slope extraction yields negative values of $c_{\rm eff}$. Such negative values are allowed within this extraction procedure. However, these negative values should not be interpreted as a physical negative central charge; rather, they are artifacts of extracting an effective slope from data in the cross-over regime that is far from either UV or IR asymptotic regimes. At larger $L/\xi$, the extracted $c_{\rm eff}$ increases and saturates to $c_{\rm eff}=0$, consistent with the area-law phase reached near the IR projective measurement fixed point. 

As an independent check, we also determine $c_{\rm eff}$ using the chord-length scaling form at fixed $L$ while varying the subsystem length $\ell$, as shown in Appendix Fig.~\ref{fig:ising_central_Charge}. This extraction yields a non-negative $c_{\rm eff}$ throughout the crossover and approaches $c_{\rm eff}=0$ in the area-law regime. The non-negativity of this chord-length estimate is consistent with the constraint imposed by strong subadditivity, which exists even for systems with disorder~\cite{Grover:2014ouz, PhysRevB.94.184202} and hence also for systems with measurement-induced randomness. We note again, as already mentioned above, that any {\it finite} loop-order epsilon expansion could not, unless resummed to all loop-orders, 
yield a vanishing value of $c_{\rm eff}$ as that obtained numerically for the infrared fixed point for the Ising case corresponding to exactly $m=3$, the lowest possible value of $m$. This, however, is not a shortcoming of the epsilon expansion itself, but rather a reflection of the fact that the expansion is not a useful tool for computing the quantity $c_{\rm eff}$ 
in cases where this quantity vanishes.

\subsection{Effective Affleck-Ludwig Boundary Entropy ($s_{\rm eff}$)}

We perform a finite-size RG crossover analysis of the Shannon entropy of the measurement record, $S_{\rm Sh}$, whose scaling form is given in Eq.~\ref{eq:boundary_entropy}. As shown in Fig.~\ref{fig:universality_ci}(b), the extracted effective boundary entropy $s_{\text{eff}}$ exhibits a monotonic decrease along the crossover, consistent with the $g$-effective theorem of Ref.~\cite{PatilLudwig20251}.
We obtain $s_{\rm eff}$ from the finite-size instantaneous slope of the Shannon entropy density $S_{\rm Sh}/L$ as a function of $1/L$. The extraction procedure is illustrated in the inset of Fig.~\ref{fig:universality_ci}(b) for the projective-limit data.

The RG crossover shows that $s_{\rm eff}$ flows from the trivial unmeasured value $s_{\rm eff}=0$, corresponding to the absence of a measurement defect in the UV regime, to the infrared value $s_{\rm eff}=-0.257(3)$. This saturated value is consistent, within error bars, with the result of Ref.~\cite{StephanMisguichPasquier}. It is also in reasonable 
agreement with the perturbative RG prediction to order $\epsilon^3$, quoted in Table~\ref{tab:table_ci}.

\subsection{Multifractality}

We investigate
multifractal scaling in the weakly measured critical Ising ground state using moments of the Born-averaged connected spin-spin correlation function, $\overline{\left|\langle\delta\sigma_i^z \delta\sigma_j^z\rangle_{\vec{m}}\right|^N}$, computed in the post-measurement ensemble with periodic boundary conditions. For finite measurement strength $\lambda$, these moments exhibit the power-law scaling form of 
Eq.~\ref{eq:tci_correlation}, as demonstrated in Appendix Fig.~\ref{fig:ising_multifractality}.

The exponents $X_{1}^{(\sigma)}$ and $X_{2}^{(\sigma)}$, extracted from log-log fits of the first and second moments, show finite-size RG crossover collapse as functions of $L/\xi$ and approach saturated infrared values, as illustrated in Fig.~\ref{fig:universality_ci}(c). In particular, the infrared exponents show a small but systematic deviation from the simple linear relation $X_{2}^{(\sigma)}=2X_{1}^{(\sigma)} = 1/4$ in the unmeasured UV regime, instead satisfying the multifractal convexity condition $X_{2}^{(\sigma)} < 2X_{1}^{(\sigma)}$. 
Although this deviation is numerically modest, it suggests that the post-measurement critical Ising ensemble possibly develops a multifractal structure analogous to that found for tricritical Ising in the previous section. 
The saturated infrared values of $X_{1}^{(\sigma)}$ and $X_{2}^{(\sigma)}$ are reported in Table~\ref{tab:table_ci}.

Unlike the TCI energy-measurement case, the Ising spin-measurement protocol does not produce a staggered odd-even structure in the correlation data. However, a microscopic effect appears in the projective limit $\lambda=1$. In this limit, the measurement projects each realization into the $\sigma^z$ basis, so the connected correlator $\langle\delta\sigma_i^z\delta\sigma_j^z\rangle_{\vec{m}}$ vanishes trivially in every measured state. This projective-limit vanishing is a consequence of measuring the same local operator whose connected correlator is subsequently evaluated, and should be distinguished from the universal scaling behavior observed at finite measurement strength.

\begin{table}[h!]
\caption{\label{tab:table_ci}Universal exponents obtained using numerical and controlled perturbative RG epsilon expansion approaches for ground state of critical Ising subjected to weak spin measurement.}

\begin{tabular}{ 
|p{2.0cm}||p{1.1cm}|p{1.7cm}|p{1.7cm}|p{1.5cm}| }
 \hline
 \multicolumn{1}{|c||}{Critical Ising} & \multicolumn{4}{c||}{Exponents/Universal numbers} \\
 \hline
 \multicolumn{1}{|c||}{}& $c_{\rm{eff}}$ & $s_{\rm{eff}}$ & $X_{1}^{(\sigma)}$ & $X_{2}^{(\sigma)}$ \\
 \hline
 Numerics & 0 & -0.257(3) & 1.14(5) & 2.01(6) \\
$\epsilon$-expansion
 & $<0$  & -0.17 {\tiny \hspace{0.7cm}(1-loop)} & 1.4375 {\tiny \hspace{0.5cm}(2-loop)} & 1.1875 {\tiny \hspace{0.5cm}(2-loop)} \\  
 \hline
\end{tabular}

\end{table}

\section{Conclusions}
We numerically investigated the critical properties of post-measurement ensembles obtained by applying weak measurements to critical ground states. Our study is motivated by the RG 
analysis
of Ref.~\cite{PatilLudwig2024}, which considered cases where the
intrinsic randomness from weak measurements 
acts 
as a relevant perturbation about the unmeasured critical fixed point and drive the post-measurement ensemble toward a measurement-dominated fixed point whose universal properties can be studied perturbatively in an epsilon expansion about the clean unmeasured fixed point.
We focused on two representative examples: the tricritical Ising ground state subjected to weak energy measurements and the critical Ising ground state subjected to weak spin measurements.

Using extensive finite-size crossover analyses, we characterized the universal properties of these post-measurement ensembles. We extracted the effective entanglement central charge $c_{\rm eff}$ from the measurement-averaged entanglement entropy, the effective Affleck-Ludwig boundary entropy $s_{\rm eff}$ from the Shannon entropy of measurement 
record, and multifractal scaling exponents from
Born-averaged moments of 
connected correlation functions of the post-measurement ensemble. The resulting universal quantities and exponents are summarized in Tables~\ref{tab:table_tci} and \ref{tab:table_ci}. Their 
qualitative
agreement with perturbative RG predictions, where available, provides a clean numerical test of the universal critical properties predicted by the epsilon expansion analysis of the
replicated field theory description.

For the tricritical Ising case, corresponding to the $m=4$
conformal unitary minimal model, our results support an RG crossover from the unmeasured tricritical Ising fixed point to a 
weak-measurement
infrared fixed point. This fixed point
has logarithmic measurement-averaged entanglement, which
is characterized by a nonzero entanglement effective central charge $c_{\rm eff}$, a universal effective boundary entropy $s_{\rm eff}$, and multifractal correlation-function exponents. In contrast, for the critical Ising case, corresponding to the $m=3$ unitary minimal model, we 
\textit{do not} find a weak-measurement
fixed point, and rather
find that the post-measurement ensemble flows under RG
directly toward the projective measurement fixed point,
where the measurement-averaged entanglement obeys an area law in the thermodynamic limit. 

Together, these results clarify the RG-flow diagram of weakly measured multicritical Ising ground states for the considered measurement protocols. The distinct behavior of the $m=3$ and $m=4$ cases supports the scenario illustrated in Fig.~\ref{fig:model_setup}(c), where finite $m>3$ minimal models are expected to exhibit a weak-measurement
fixed point at a
finite 
effective measurement strength $\Delta$, while in the critical Ising case, i.e. at $m=3$,
the weak-measurement and the projective-measurement fixed point become one and the same, and
the RG flow directly takes the system to the
projective limit. 
Heuristically, this direct RG flow to the projective limit in the critical Ising case
can be understood by noting that the measurement operator for all problems in Fig.~\ref{fig:model_setup} (c) with odd $m>3$, which formed the basis of epsilon expansion in Ref.~\cite{PatilLudwig2024}, is a \textit{subleading} spin field in the unmeasured multicritical Ising theory.
However, only for the case of lowest possible $m$, $m=3$ corresponding to the critical Ising case, the measurement operator actually turns out to be the leading
spin operator from the unmeasured CFT ground state.
Since we are then  performing weak measurements with the
order-parameter, the most-relevant field in the Ising CFT, the
observed flow directly to the projective limit is a reasonable RG flow.
We note that the perturbative epsilon expansion~\cite{PatilLudwig2024} to any {\it finite} order in $\epsilon$ is, by construction (unless resummed to all loop-orders), unable to capture the projective limit feature and the area law entanglement entropy ($c_{\text{eff}}=0$) of the infrared fixed point for the Ising case, corresponding to $m = 3$. This is not a shortcoming of the expansion itself. And, following the analytic continuation in $m$ suggested by Fig.~\ref{fig:model_setup} (c), it is well conceivable that the $\epsilon$-expansion results~\cite{PatilLudwig2024} at $m = 3$ would capture non-trivial critical properties such as the effective boundary entropy $s_{\text{eff}}$ and the power-law exponents for the moments of the correlation functions~\footnote{ upon possibly, if necessary, going to sufficiently high loop order (and resummation)} of the post-measurement ensemble for the critical Ising case~\cite{Note8}.

This provides a concrete example of how intrinsic randomness from quantum measurements can generate universal long-distance behavior in post-measurement ensembles of critical ground states, while still allowing a controlled characterization through both analytical RG and numerical finite-size RG crossover analyses.

{\it Acknowledgments -- } This work was partly supported by the Swiss National Science Foundation (R.V., grant 10008234),  the Foundation for the University of Geneva (R.V.), and by the US Department of Energy, Office of Science, Basic Energy Sciences, under award No. DE-SC0023999 (A.K. and R.V.).  R.V., R.A.P. and A.W.W.L acknowledge the hospitality of KITP during
the 2025 program ``Learning the Fine Structure of Quantum Dynamics'', supported by NSF PHY-2309135, as well as the Nordita 2026 program ``Entanglement Dynamics: Open Quantum Systems, Monitored Circuits, and Topological Order'', where part of this work was done.
 We thank Ehud Altman, Sam Garratt, Kabir Khanna, Sara Murciano, Pablo Sala,  and Sarang Gopalakrishnan for insightful discussions and/or collaborations on related topics.

\appendix

\section{Entanglement Entropy Scaling and Effective Central Charge ($c_{\rm eff}$)}

\begin{figure}[t]
	\centering
	\includegraphics[width=0.5\textwidth]{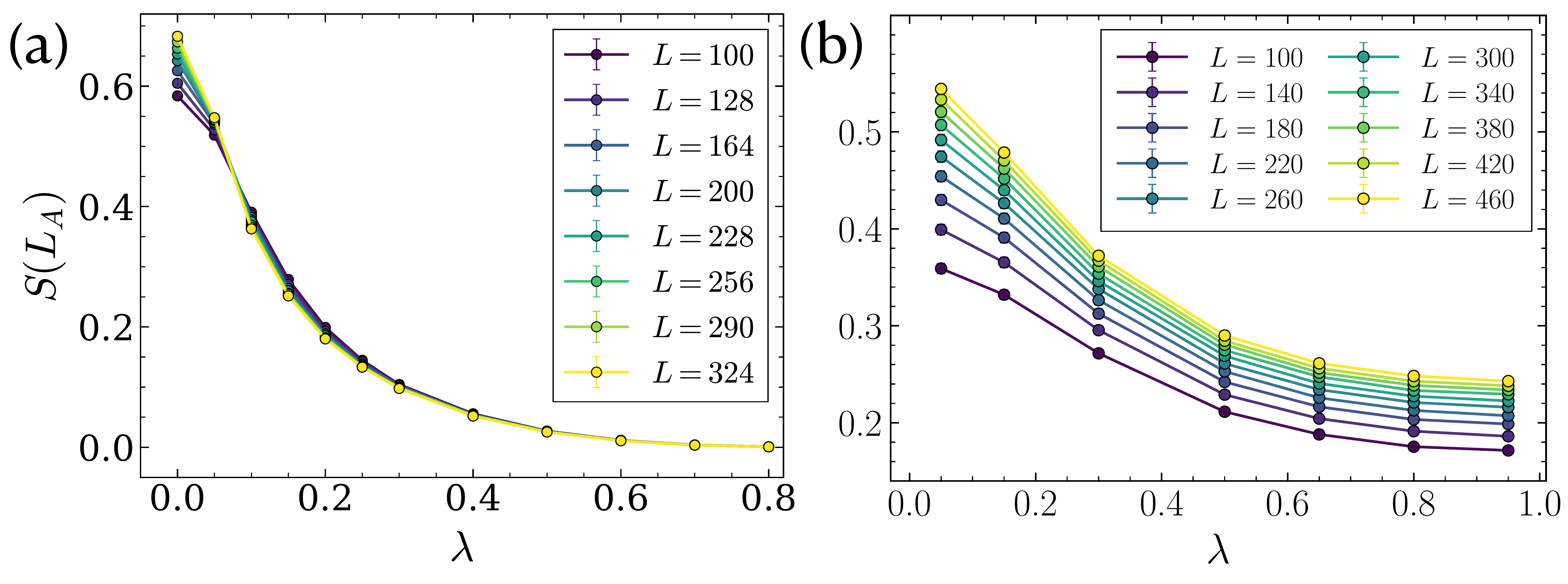}
    \caption{\textbf{Scaling of Entanglement entropy.} We perform finite-size scaling of bipartite entanglement entropy $S(L_A)$ between two halves of the system for the ground state of (a) critical Ising and (b) tricritical Ising subjected to their respective measurement protocols under open boundary condition. (a) The ground state of critical Ising with $\sigma_z$ weak measurement on each site leads to an area law entanglement at any non-zero measurement strength. For intermediate measurement strength regime and finite-size $L$, the entanglement entropy decreases with increase in system size $L$. This region  extends toward $\lambda \to 0$ on further increasing $L$ which stays consistent with the crossover scaling collapse observed in Fig.~\ref{fig:universality_ci}(a). However, (b) tricritical Ising with local energy measurement on alternating links scales logarithmically with $L$ at all measurement strength $\lambda$.      
    }
	\label{fig:entanglement_scaling}
\end{figure}

\begin{figure}[t]
	\centering
	\includegraphics[width=0.5\textwidth]{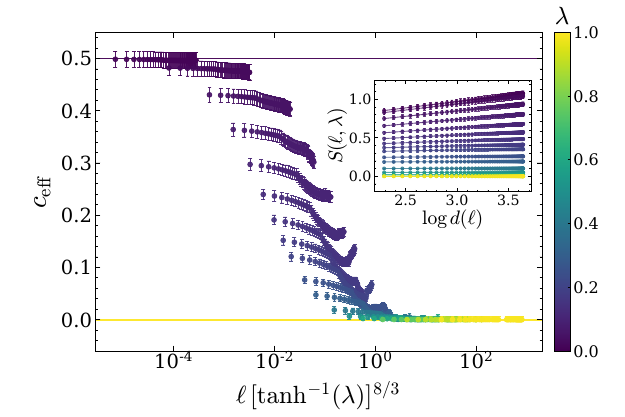}
    \caption{
    \textbf{Ising effective central charge ($c_{\rm eff}$).}
    We plot the effective central charge obtained from bipartite entanglement entropy between contiguous regions by varying the subsystem ($A$) length $\ell$, for a fixed system size $L = 120$ under periodic boundary condition. In the inset, we plot the entanglement entropy against the $\log$ of chord length $d(\ell)$ where each data point is averaged over $2\times 10^5$ realizations. At small measurement strength $\lambda$ and finite L, we get a linear trend with $\log(d(\ell))$ that becomes independent of $d(\ell)$ at large $\lambda$. The obtained $c_{\rm eff}$  from the slope of the inset plot satisfies the non-negative constraint $c_{\rm {eff}|_{L}} \geq 0$ at all $\lambda$, in accordance with the strong subadditivity property. Moreover, the flow of $c_{\rm eff}$ against $\ell/\xi$, where  $\xi = \tanh^{-1}(\lambda)^{-8/3}$  corroborates the value of ultraviolet and infrared fixed point shown using their respective reference lines at their corresponding $\lambda$ points. In this procedure, we probe both the subsystem scale $d(\ell)$ and the global finite-size scale $L$ which can lead to residual dependence on $L/\xi$, so a single-parameter collapse in $\ell/\xi$ or $d(\ell)/\xi$ is not expected to be sharp.
    Though to make the data smooth we perform moving average of the slope over a fit window of 4 data points.}
	\label{fig:ising_central_Charge}
\end{figure}

\begin{figure}[t]
	\centering
	\includegraphics[width=0.5\textwidth]{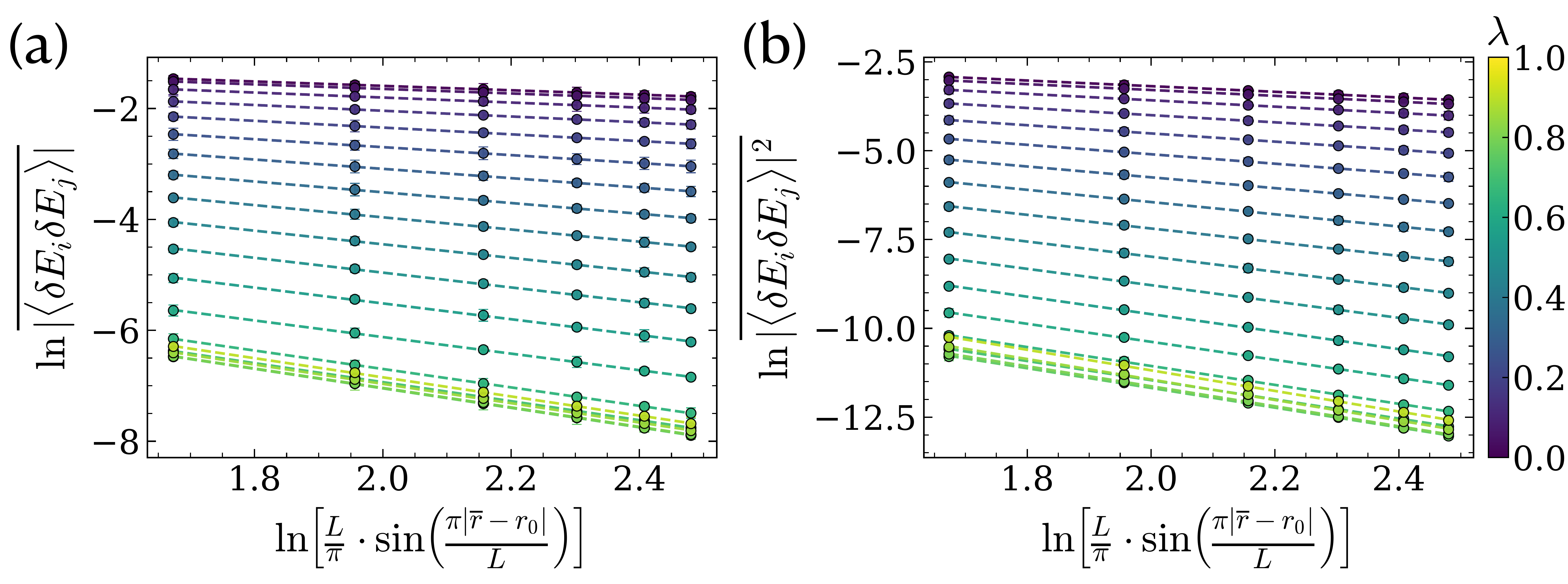}
    \caption{
    \textbf{Tricritical Ising multifractality.} We confirm the power-law trend of measurement averaged moments of connected energy correlation function of the post-measured  ground state of tricritical Ising with periodic boundary condition. (a) The first moment of the measurement averaged connected energy $E_i$ two-point correlation function supports power law scaling at all measurement strength $\lambda \in [0,1)$, shown using a linear trend with a log-log plot of $\overline{|\langle \delta E_i \delta E_j\rangle_{\vec{m}}|}$ against $\frac{L}{\pi}\cdot \sin(\frac{\pi|\bar{r}-r_0|}{L})$. (b) The second moment again point to power law trend for the same set of measurement strength but with an independent exponent. The data is shown for $L=40$ and each data point is averaged over $10^5$ realizations.  
    }
	\label{fig:tci_multifractality}
\end{figure}

\begin{figure}[t]
	\centering
	\includegraphics[width=0.5\textwidth]{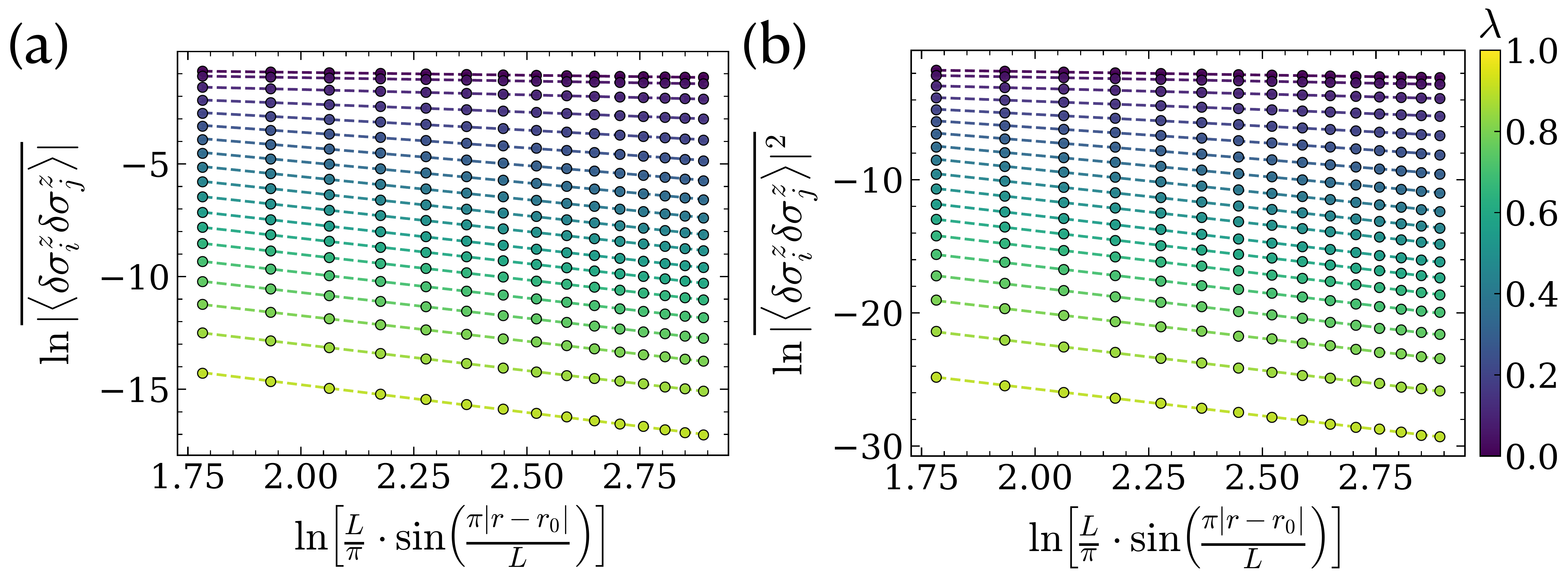}
    \caption{
    \textbf{Critical Ising multifractality.} We confirm the power-law trend of measurement averaged moments of connected spin correlation function of the post-measured GS of CI with periodic boundary condition. (a) The first moment of the measurement averaged connected $\sigma_i^z$ two-point correlation function supports power law scaling at all measurement strength $\lambda \in [0,1)$, shown using a linear trend with a log-log plot of $\overline{|\langle \delta\sigma_i^z \delta\sigma_j^z\rangle_{\vec{m}}|}$ against $\frac{L}{\pi}\cdot \sin(\frac{\pi|r-r_0|}{L})$. (b) The second moment again point to power law trend for the same set of measurement strength but with an independent exponent. The data is shown for $L=80$ and each data point is averaged over $10^5$ realizations.  
    }
	\label{fig:ising_multifractality}
\end{figure}

For a finite one-dimensional critical ground state described by a translationally invariant $(1+1)$D CFT, the bipartite von Neumann entanglement entropy obeys a universal logarithmic scaling form. For a ground state of length $L$ with periodic boundary conditions, the entanglement entropy of a subsystem of length $\ell$ is

\begin{equation}
\overline{S(\ell)}=\frac{c}{3}\log d(\ell)+\mathcal{O}(1),
\label{eq:entanglement_chord_formula}
\end{equation}

where $d(\ell)=\frac{L}{\pi}\sin\left(\frac{\pi \ell}{L}\right)$ is the chord length. The coefficient of the logarithm is fixed by the central charge of the underlying CFT. For the critical Ising and tricritical Ising ground states, these values are $c=1/2$ and $c=7/10$, respectively, obtained from the $m=3$ and $m=4$ unitary minimal models.

In the post-measurement ensemble, the central charge is replaced by an
entanglement effective central charge $c_{\rm eff}$, which characterizes the logarithmic part of the measurement-averaged entanglement entropy. We use two complementary procedures to extract this quantity. In the first method, we fix the subsystem fraction $\ell/L=1/2$ and vary the total system size $L$. This is the extraction used in Figs.~\ref{fig:universality_tci}(a) and \ref{fig:universality_ci}(a), where $c_{\rm eff}$ is obtained from the slope of $\overline{S(L/2)}$ as a function of $\log L$. For the critical Ising state under spin measurements, the finite-size data show logarithmic scaling at weak measurement strength, followed by a crossover to area-law behavior at larger $\lambda$, as shown in Fig.~\ref{fig:entanglement_scaling}(a) and in the inset of Fig.~\ref{fig:universality_ci}(a). At sufficiently large $\lambda$, the entanglement becomes independent of $L$, giving $c_{\rm eff}=0$. Before reaching this asymptotic area-law regime, however, the finite-size slope can become negative. This negative value is not interpreted as a physical central charge; it reflects the use of a local finite-size slope in the crossover regime. The RG crossover collapse organizes this behavior as a function of $L/\xi$: for any fixed finite $\lambda$, the thermodynamic limit corresponds to $L/\xi\to\infty$, where $c_{\rm eff}$ approaches zero. Thus, the apparent negative-$c_{\rm eff}$ region is a finite-size crossover feature and does not survive as an asymptotic thermodynamic regime in the half-system entanglement scaling shown in Fig.~\ref{fig:entanglement_scaling}(a). In contrast, for the tricritical Ising state under energy measurements, the entanglement remains logarithmic for all measurement strengths studied, as shown in Fig.~\ref{fig:entanglement_scaling}(b), and the crossover collapse in Fig.~\ref{fig:universality_tci}(a) approaches a nonzero infrared value of $c_{\rm eff}$.

In the second method, we extract $c_{\rm eff}$ directly from the slope of entanglement as a function of logarithm of the chord length $d(\ell)=\frac{L}{\pi}\sin\left(\frac{\pi \ell}{L}\right)$. Here the total system size $L$ and measurement strength $\lambda$ are fixed, while the subsystem length $\ell$ is varied in the range $1\leq \ell\leq L/2$, using the complementarity relation $\overline{S(\ell)}=\overline{S(L-\ell)}$. This procedure probes the scale dependence of the entanglement within a single finite system and provides an independent check of the half-system extraction. For translationally invariant averaged states, the corresponding chord-length estimation of $c_{\rm eff}$ is constrained by strong subadditivity to be non-negative~\cite{Grover:2014ouz}. In Fig.~\ref{fig:ising_central_Charge}, this chord-length extraction shows a crossover from the Ising value $c=1/2$ at short distances to $c_{\rm eff}=0$ in the infrared
area-law regime. The collapse is organized by comparing the subsystem length scale to the RG crossover length. Since this procedure probes both the subsystem scale $d(\ell)$ and the global finite-size scale $L$, residual dependence on $L/\xi$ can remain, so a single-parameter collapse in $d(\ell)/\xi$ is not expected to be as sharp as the half-system collapse in $L/\xi$.

\section{Multifractal Scaling Under Weak Measurement}

Multifractal scaling is characterized by a hierarchy of independent scaling exponents associated with different moments of a correlation function. We test this behavior for the tricritical Ising and critical Ising ground states subjected to weak measurements at finite measurement strength, using periodic boundary conditions. Specifically, we analyze the first and second moments of the Born-averaged connected correlation functions in the post-measurement ensemble: the energy-energy correlator for tricritical Ising and the spin-spin correlator for critical Ising.

To verify the power-law scaling, we perform log-log linear fits of the corresponding moments as functions of the chord length for fixed system size $L$. In Fig.~\ref{fig:tci_multifractality}, we show representative fits for the tricritical Ising case at $L=40$, while Fig.~\ref{fig:ising_multifractality} shows the corresponding critical Ising fits at $L=80$. For the tricritical Ising ground state under energy measurements on alternate bonds, we average correlators for neighboring odd and even sites
to suppress the staggered lattice component that becomes pronounced at large $\lambda$; this averaged coordinate is denoted by $\overline{r}$. In contrast, the critical Ising spin-measurement protocol does not show such staggering at finite $\lambda$, and the correlation functions exhibit clean power-law behavior except in the projective limit, where the connected spin correlator vanishes trivially due to projection into the $\sigma^z$ basis.

\bibliography{Ref}
\end{document}